\documentclass{PoS}
\usepackage{amsmath}
\usepackage[utf8x]{inputenc}

\newcommand{\eq}{\begin{equation}}
\newcommand{\eqx}{\end{equation}}
\newcommand{\eqn}{\begin{eqnarray}}
\newcommand{\eqnx}{\end{eqnarray}}
\newcommand{\alg}{\begin{align}}
\newcommand{\algx}{\end{align}}

\newcommand{\f}[2]{\frac{#1}{#2}}

\newcommand{\lm}{\lambda}
\newcommand{\Lm}{\Lambda}

\newcommand{\dl}{\delta}
\newcommand{\Dl}{\Delta}
\newcommand{\bt}{\beta}

\newcommand{\al}{\alpha}
\newcommand{\eps}{\varepsilon}
\newcommand{\nn}{{\cal N}}

\newcommand{\OO}[1]{{\cal O}\left(#1\right)}
\newcommand{\cor}[1]{\left\langle#1\right\rangle}

\newcommand{\teff}{T_{eff}}

\title{AdS/CFT and applications}

\ShortTitle{AdS/CFT and applications}

\author{\speaker{Romuald A. JANIK}%
\\
        Jagiellonian University, Kraków\\
        E-mail: \email{romuald@th.if.uj.edu.pl}}

\abstract{In this talk I review some selected applications
of the AdS/CFT correspondence to the study of nonequillibrium
dynamics of strongly coupled plasma. I briefly describe the
AdS/CFT methods and five recent studies dealing with
various features of plasma evolution, in particular
the transition to hydrodynamics, thermalization, 
very high order hydrodynamics, the
dynamics of shock wave collisions and preequilibrium radial flow.}

\FullConference{The European Physical Society Conference on High Energy Physics \\
                 18-24 July, 2013\\
                 Stockholm, Sweden}

\begin{document}

\section{Introduction}

The AdS/CFT correspondence \cite{ADSCFT} is the postulated equivalence between 
a four-dimensional gauge theory
-- the maximally supersymmetric $\nn=4$ Super-Yang-Mills theory and superstring theory
on $AdS_5 \times S^5$, a specific ten-dimensional curved background.
Despite the fact that there is no general proof of this correspondence, currently we
have overwhelming evidence that it is indeed true, including very explicit
comparisions between standard gauge theory perturbative computations at quite high loop
order and string computations on the dual side.

One reason why the AdS/CFT correspondence is a very valuable tool for the understanding
of gauge theory dynamics is the fact that in the nonperturbative strongly coupled
regime, the dual string description simplifies dramatically and reduces
either to classical gravity or (semi-)classical strings depending on
concrete gauge theory observables. Thus it provides us with completely
new ways of looking at nonperturbative gauge theory physics.
In particular, many dynamical problems in gauge theory dynamics are
translated into problems in higher dimensional General Relativity.
Apart from the purely pragmatic interest in obtaining gauge theory
answers in this way, the very possibility of doing so is fascinating
theoretically as is the link between so apparently diverse
branches of physics as General Relativity and gauge theory. 

Currently, the range of applications of the AdS/CFT correspondence is
immense -- the original article \cite{ADSCFT} had, as of $15^{th}$ November 2013, 
9389 citations. So in this talk I will concentrate only on reviewing some of its recent
applications to the study of the dynamics of strongly coupled plasma.
Also, due to time and space constraints, I will pick out just one
strand of research concentrating on nonequilibrium dynamics of plasma
and the transition to a hydrodynamic behaviour.

The applications of the AdS/CFT correspondence in this domain
essentially started with the calculation of the shear viscosity
of strongly coupled plasma in the $\nn=4$ gauge theory \cite{PSS}. 
A very valuable insight came from the realization that the
ratio of the shear viscosity to entropy density is universal and equal to $1/4\pi$
for a very wide class of \emph{strongly coupled} gauge theories \cite{KSS}.
A feature worth emphasizing is that even if the corresponding value 
for realistic QCD quark-gluon plasma is not exactly the same,
the value of $1/4\pi$ from the AdS/CFT correspondence gave 
a ballpark figure as a point of reference for what would
be expected for QCD plasma.
I believe that this kind of input from the AdS/CFT correspondence
may be very useful. This is especially important for nonequilibrium 
phenomena for which effective nonperturbative methods are quite scarce.

The preceeding investigations involved essentially the emergence of
linearized hydrodynamics around a static uniform plasma system.
In \cite{JP1} it was shown that the AdS/CFT correspondence
predicts also fully nonlinear hydrodynamic flow (Bjorken flow)
at late times in a boost-invariant setting.
Since then, the emergence from AdS/CFT of nonlinear viscous hydrodynamics
without any symmetry assumptions was understood \cite{FLUIDGRAVITY}
as arising from a gradient expansion within the gravitational Einstein's equations.

Once we approach truly out of equilibrium processes, the gradients
become large and an analytical scheme for solving Einstein's equations
is no longer feasible in practice. The first work on applying numerical
methods in this context was \cite{CY1} dealing with isotropization
of a spatially uniform plasma. All the investigations which I describe
in more detail in this talk involve numerical relativity methods.

The plan of this talk is as follows. I will start with a discussion of
the similarities and differences between $\nn=4$ and QCD plasma, then I
will review briefly the AdS/CFT methods used for the description
of evolving plasma systems and mention some key questions.
In the rest of the talk I will review five selected recent studies
of various aspects of out-of-equilibrium plasma dynamics using
AdS/CFT. I will close the talk with some conclusions.
The topics covered in this talk are neccessarily just a selection. 
For a more in-depth recent review see \cite{GUBSERREVIEW}.

\section{$\nn=4$ plasma versus QCD plasma}

The technically simplest case of the AdS/CFT correspondence is its original form
stating a duality between the supersymmetric $\nn=4$ gauge theory and 
superstring theory on the curved ten-dimensional $AdS_5 \times S^5$ background.
The recent investigations of the nonequilibrium dynamics of strongly coupled
plasma were performed, almost exclusively, within this context. This leads us
at once to the question, to what extent will such studies be relevant for the case
of realistic QCD quark-gluon plasma.

Of course, the $\nn=4$ gauge theory is completely different from QCD in the vacuum.
It is supersymmetric, has no mass scale, being exactly conformal also on the quantum 
level. Consequently it has no confinement and exhibits an exactly Coulombic potential
between colour sources. However once we consider the theory at nonzero temperature,
or more generally for a state with some nonzero energy density of the order of $N_c^2$,
we get significantly more qualitative similarities. Supersymmetry is 
explicitly broken through the introduction of temperature, which becomes the dominant
scale in the problem. The theory is still deconfined but, in case of AdS/CFT studies,
strongly coupled. But now this is not far from the qualitative behaviour of the QCD quark-gluon 
plasma above $T_c$, which is also deconfined and strongly coupled.
In fact even at weak coupling, the differences between $\nn=4$ plasma and QCD plasma
amount essentially just to differences in the number of degrees of freedom \cite{MROW}.

Given the above, one should, however, be aware of certain differences, which may be or 
need not be relevant for the physical phenomena studied. Firstly, the $\nn=4$ gauge theory
has no running coupling, so even at very high temperatures the coupling remains strong.
This may not be a problem for the description of plasma \emph{per se} as
the asymptotic regime for temperatures is certainly not reached, it may however be 
very relevant for mixed perturbative/nonperturbative processes e.g. hard probes.
Secondly, the equation of state for the $\nn=4$ plasma remains exactly conformal ($E=3p$). This is certainly
not true for QCD plasma close to the phase transition/crossover, but may be 
quite a good approximation for a range of higher temperatures. 
Finally, in the $\nn=4$ theory, there is no confinement/deconfinement phase transition
so the $\nn=4$ plasma expands and cools indefinitely. Thus, one cannot use this theory
to model behaviour in the vicinity of the phase transition.

One should note that the above differences are so acute just for the simplest case of the
AdS/CFT correspondence. There are explicit, albeit more complicated, examples of
AdS/CFT description of theories for which some of these differences are lifted.

To summarize, the applicability of using $\nn=4$ plasma to model real world
phenomenae depends on the physical problems considered. The basic motivation
for undertaking this study is that one can use it as a theoretical laboratory where we may compute from ‘first principles’ nonequilibrium nonperturbative dynamics
all the way from some nonequilibrium initial state to a precise hydrodynamic
description. All this can be done within the same calculational framework. 
We may thus gain qualitative 
insight into the physics which is very difficult to access using other methods.
Then we may use the behaviour of strongly coupled $\nn=4$ plasma
as a point of reference for analyzing/describing QCD plasma
and eventually consider more realistic theories with AdS/CFT duals.

\section{The AdS/CFT description}

In this section we will review the methods used for the AdS/CFT description of an evolving
strongly coupled plasma system. Since the plasma system is a complex state involving $\OO{N_c^2}$
degrees of freedom, it is described through an effective 5D geometry\footnote{This geometry
can be uplifted to a 10D string background in a canonical way. However this will not play 
any role in the following.}
\eq
\label{e.geom}
ds^2=\f{g_{\mu\nu}(x^\rho,z)\, dx^\mu dx^\nu +dz^2}{z^2} \equiv g^{5D}_{\al\bt} dx^\al dx^\bt
\eqx
where $x^\mu$ are the boundary (Minkowski) coordinates, while $z \geq 0$ is the $5^{th}$
coordinate which extends from the 4D boundary at $z=0$ into the 5D bulk.
Let us note that this simple description arises only at strong coupling. Then there is a 
large mass-gap (going to infinity as $\lm\to \infty$) between the gravity modes and massive string excitations in $AdS_5 \times S^5$, so one can reduce the dynamics to just the gravity sector.
In contrast, for weakly coupled $\nn=4$ plasma, there would be no justification for
neglecting the massive string modes and a dual string description is currently beyond our reach.

\begin{figure}[t]
\centerline{\includegraphics[height=5cm]{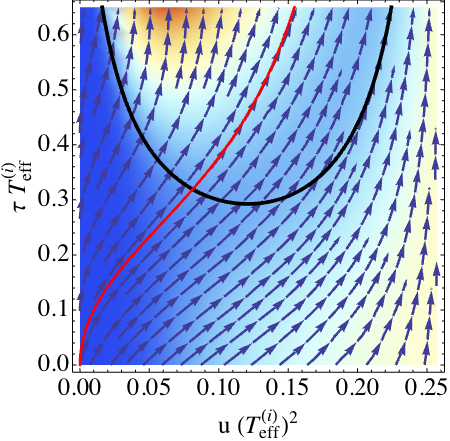}}
\caption{The procedure for extracting the entropy from a boost invariant numerical
simulation. The entropy at a given time $\tau$ (here $\tau=0$) is read off from the area element
of the apparent horizon at the point of intersection with a null
geodesic emitted from the boundary at time $\tau$.}
\label{fighor}
\end{figure}

The time evolution of the plasma system follows from 5D Einstein's equations with negative
cosmological constant
\eq
\label{e.einst}
R_{\al\bt}-\f{1}{2}g^{5D}_{\al\bt} R - 6\, g^{5D}_{\al\bt}=0
\eqx
Once the background geometry is known, either analytically or numerically,
physical observables may be extracted. In the present context, the most
relevant observable is the expectation value of the energy momentum tensor 
$\cor{T_{\mu\nu}(x^\rho)}$ which can be read off from the asymptotic behaviour of the geometry
(\ref{e.geom}) near the boundary $z=0$. Namely
\eq
g_{\mu\nu}(x^\rho,z) =\eta_{\mu\nu}+g^{(4)}_{\mu\nu}(x^\rho)\, z^4+ \OO{z^6}
\quad\quad
\cor{T_{\mu\nu}(x^\rho)}= \f{N_c^2}{2\pi^2}\, g^{(4)}_{\mu\nu}(x^\rho)
\eqx
Once the energy-momentum tensor is known, one can check when it starts to have hydrodynamic
form to investigate the transition to hydrodynamics, whether in a local frame it is isotropic
in order to check for local thermal equilibrium etc.

A second important physical observable which may be extracted from the geometry is
a local entropy density. The prescription, introduced in \cite{ENTROPY1} and refined later, 
is more complex (it can be obtained from an area element of an apparent horizon transported to the boundary 
through a null geodesic (see fig.~\ref{fighor}))
and suffers from some ambiguities away from equilibrium. Yet in the hydrodynamic
regime it has all the right properties of phenomenological hydrodynamic entropy density.

\section{Key questions}

A key question relevant for heavy-ion collisions at RHIC and LHC is
the time of applicability of a hydrodynamic description. Most of the studies described
in the present talk address this problem in the context of the dynamics
of strongly coupled $\nn=4$ plasma which is initially in a far from equillibrium state.

We are also
interested in more refined characteristics of the transition to hydrodynamics. In particular, we would like to know whether
at the transition to hydrodynamics we have already local thermal equilibrium (in the sense
of isotropy of the energy-momentum tensor).
In addition we would like to understand
what characterization of the initial state determines the main features of its transition
to hydrodynamics. Furthermore we would like to have the same kind of detailed understanding
of a collision process and to what extent is Bjorken boost-invariant flow recovered dynamically.
Finally, it is very important to understand what are the implications of AdS/CFT
for the case incorporating radial (and eventually elliptic) flow.

\section{Five recent studies}

In the remaining part of the talk I will present five selected recent studies of
various features of the dynamics of strongly coupled plasma. They progressively cover
larger pieces of a generic collision process, starting from searching for
common features in the equilibration of boost-invariant plasma from diverse
initial conditions~\cite{US1,US2}, through a very detailed study of the general properties 
of hydrodynamic derivative expansion~\cite{USBOREL}, the incorporation of radial 
dynamics~\cite{WilkeRadial},
the treatment of shock-wave collisions~\cite{MHSHOCK} and finally a hybrid AdS/CFT/realistic
hydro+cascade model of a heavy-ion collision~\cite{WRS}.  

\subsubsection*{Study I. Evolution of boost-invariant plasma from various initial
conditions at $\tau= 0$}

In the papers \cite{US1,US2} we have examined the evolution of a boost-invariant plasma system
(with no transverse dependence) from various initial conditions at $\tau=0$.
The initial conditions were represented by different initial geometries used
as initial data for solving Einstein's equations~(\ref{e.einst}).
A~weak coupling analog of this setup would be to start from a gas of 
gluons with various \emph{nonthermal} momentum distributions at some initial time
and then following its evolution up to a hydrodynamic flow and thermalization.
The chief motivation of this study, w.r.t. earlier numerical work \cite{CYBOOST},
was to investigate very general initial conditions right at $\tau=0$ and
seek regularities in the far from equilibrium behaviour.

The key questions addressed in this study were i) when does hydrodynamics become
applicable? ii) is $T_{\mu\nu}$ approximately isotropic at the transition
to hydrodynamics and iii) is there some simple physical characterization
of the initial state which would determine the characteristics of its
transition to hydrodynamics. This last question is especially acute as
clearly \emph{a-priori} the space of initial configurations is infinite-dimensional
(i.e. parametrized by a function). In fact all the initial conditions considered
in~\cite{US1,US2} were normalized to have the same energy density at $\tau=0$,
so we seek for a different characterization.

In order to factor out the trivial difference between QCD plasma and $\nn=4$ plasma
coming from different number of degrees of freedom, it is convenient to
use instead of the energy density $\eps(\tau)$ an \emph{effective temperature} $\teff$
defined as the temperature of a thermal system with energy density
equal to the instantenous energy density $\eps(\tau)$, namely
\eq
\eps(\tau) \equiv N_c^2 \cdot \frac{3}{8} \pi^{2} \cdot \teff^4
\eqx
In addition, it is convenient to form a dimensionless quantity 
$ w \equiv \teff \cdot \tau$.\\
The basic findings of ref. \cite{US1} were as follows.

\begin{figure}[t]
\centerline{\includegraphics[height=5cm]{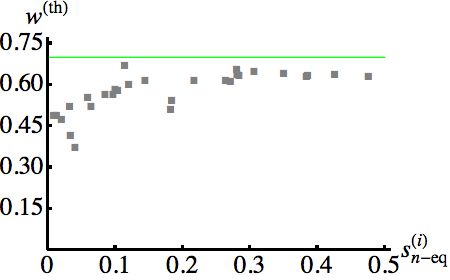}}
\caption{$w=\teff \tau$ at the transition to hydrodynamics for various initial conditions (parametrized by the initial entropy) \cite{US1}.}
\label{fig1}
\end{figure}

Firstly, for all initial conditions considered, we found that viscous hydrodynamics works to
a very high accuracy as soon as $w>0.7$ and in most cases 
even for lower $w$ (see fig.~\ref{fig1}). In order to get a feeling for this number,
e.g. quite early initial conditions for hydrodynamics at RHIC assumed in \cite{BCFK}, 
namely $\tau_0=0.25\,fm$ and $T_0=500\, MeV$ correspond to $w=0.63$.
It is very encouraging that the $\nn=4$ strongly coupled plasma gives
an answer in line with experimental expectations for realistic quark-gluon plasma.

Secondly, the plasma, when it starts being very well described by viscous hydrodynamics
is still far from being isotropic. Namely it has a sizable pressure anisotropy
\eq
\Dl p_L \equiv 1-\f{p_L}{\eps/3} \sim 0.7
\eqx
This means that the plasma is still not in local thermal equilibrium, but the
final approach to equilibrium occurs not through some exotic physics
but rather is completely described by viscous effects within hydrodynamics.
So `hydrodynamization' is distinct from thermalization.
This conclusion applies for all initial conditions considered.

\begin{figure}[h]
\hfill\includegraphics[height=5cm]{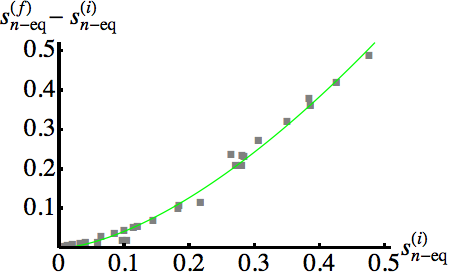}\hfill\hfill%
\includegraphics[height=5cm]{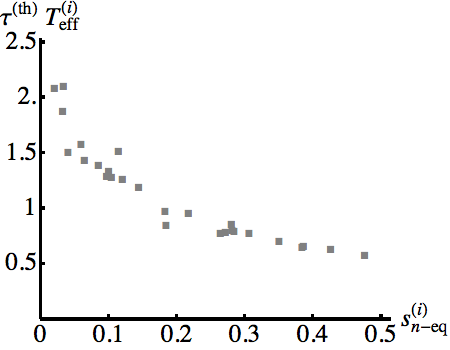}\hfill{}
\caption{Entropy production (left) and the time of transition to hydrodynamics in units of initial temperature (right) for various initial conditions parametrized by the initial entropy.
\cite{US1}}
\label{fig2}
\end{figure}

Finally, for all initial conditions, we have identified a nonzero initial entropy density
per transverse area and unit rapidity. It turns out that this single number is 
a key characterization of the initial state in the sense that there is a clear
correlation between the initial entropy and such physical observables as 
the produced entropy (see fig.~\ref{fig2}~(left))
or the time of transition to a viscous hydrodynamic
description (see fig.~\ref{fig2}~(right)).

\subsubsection*{Study II. Behaviour of high order dissipative hydrodynamics}

This study \cite{USBOREL} is more theoretical in nature, although it may also be
a point of departure for further investigations with some experimental ramifications.

The hydrodynamic description of any physical system is a quintessential example
of a derivative expansion. By including terms with more and more derivatives
in the energy-momentum tensor, we may write higher order viscous hydrodynamic
theories\footnote{This is not a purely academic endeavor, as it is known
that ordinary ($1^{st}$ order) relativistic viscous hydrodynamics 
is incompatible with causality.}.
At each order new transport coefficients appear. As with any perturbative
expansion it is of general interest to understand its convergence properties,
whether it is asymptotic, what are the singularities in the Borel plane etc.
Apart from these formal properties, a natural question arises whether one
can define a resummed all-order hydrodynamics. A proposal based on the linearized
regime of AdS/CFT has been made by Lublinsky and Shuryak in 
a very interesting work~\cite{LS} and applied phenomenologically. 
In this study, the relevant series is calculated at
the full nonlinear level.

\begin{figure}[t]
\centerline{\includegraphics[height=5cm]{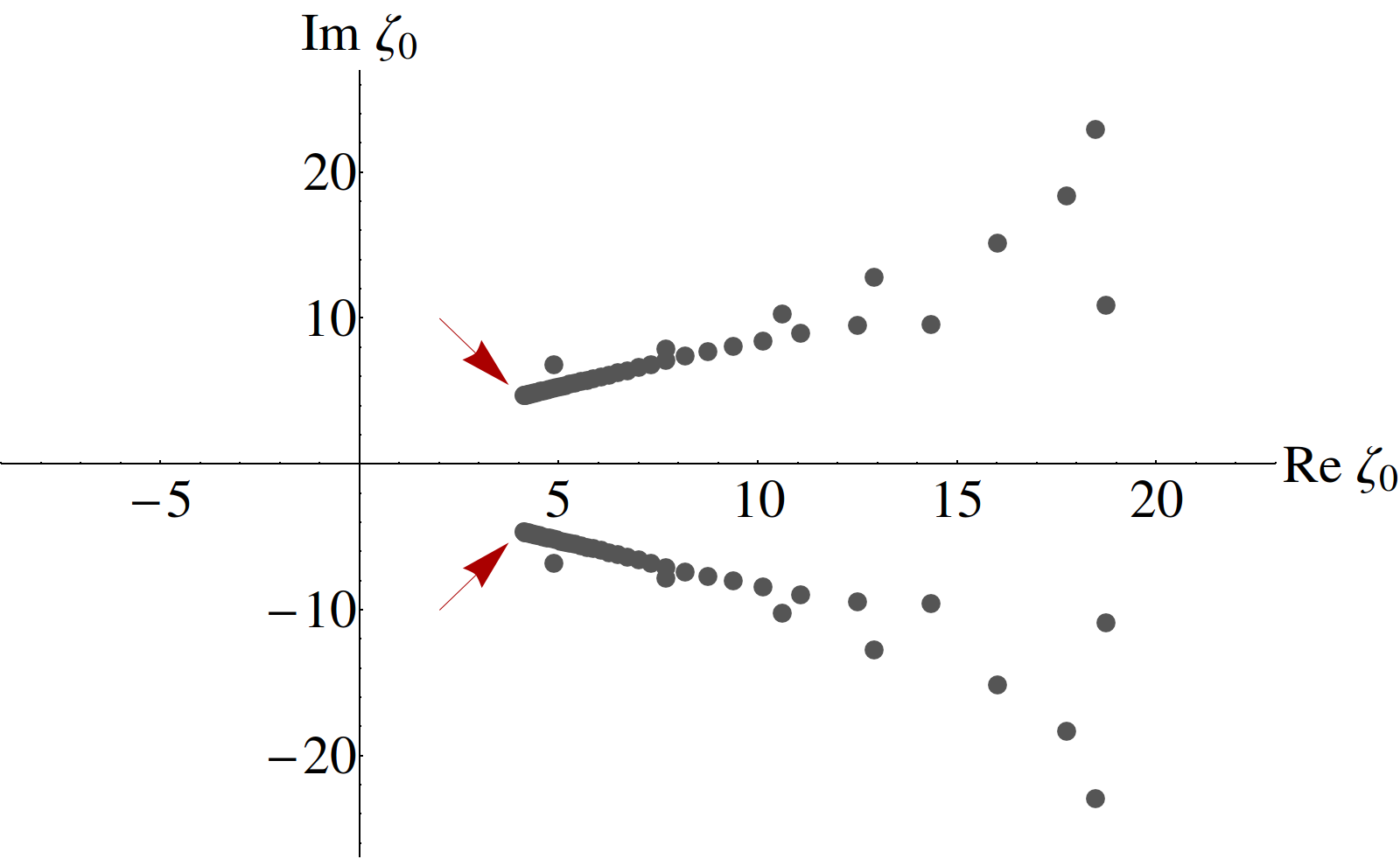}}
\caption{Singularities of the hydrodynamic gradient expansion in the Borel plane.
The cuts marked by the arrows correspond to the lowest nonhydrodynamic
excitations of the plasma (quasi-normal modes).}
\label{fig3}
\end{figure}

In \cite{USBOREL}, we studied the hydrodynamic derivative expansion in the
boost invariant kinematics. Previously just $3$ subleading terms were known
from analytical computations. Using the iterative scheme of fluid/gravity
duality we extracted $240$ terms in the series, where each order
includes contributions from new transport coefficients.

Given this many terms, we could unambigously determine that the radius of
convergence of the hydrodynamic gradient expansion is zero and that it is
an asymptotic series. The pattern of singularities on the Borel plane is shown
in fig.~\ref{fig3}. The two branch cuts closest to the origin turn out
to correspond exactly to the lowest \emph{non-hydrodynamic} degrees of freedom
(so-called quasi-normal modes). Moreover, there does not seem to be any 
poles on the positive real axis, suggesting the existence of Borel-resummed
hydrodynamics, at least in the boost-invariant context.

Such resummed hydrodynamics may be used along the lines of the investigation
from \cite{LS}, but also it may serve as a reference point for the understanding
of the lowest nonhydrodynamic excitations in the full numerical simulation.

\subsubsection*{Study III. Evolution of boost-invariant plasma with radial flow}

A natural, but technically involved, extension of the previous studies 
of boost-invariant plasma expansion is the incorporation of radial flow. 
This has been done in \cite{WilkeRadial}, where evolution from
Glauber model like initial conditions (with radius $\sim 6.5\, fm$) were studied
with an initial effective temperature adapted to a realistic value at $\tau=0.6\, fm$.
Here due to a different
numerical framework than in \cite{US1,US2} the initial conditions were setup 
at a small but nonzero
initial time $\tau_{ini}=0.12\,fm$. 

The outcome was that very good agreement with hydrodynamics was established
from $\tau=0.35\, fm$ as long as one stayed away from the edges.
Another feature characteristic of using the AdS/CFT methods to model
a part of the nonequilibrium plasma expansion is that generically
one is bound to obtain nontrivial velocity profiles at the transition
to hydrodynamics. This feature was absent in the boost-invariant study with
no transverse dynamics, as there the flow velocity was completely
fixed by symmetry. Once one allows e.g. radial flow, true
nontrivial preequilibrium flow may occur. 
This may have important experimental consequences as will be described below.

\subsubsection*{Study IV. Thick and thin shock wave collisions}

The preceeding investigations dealt with the evolution of the plasma starting
from some initial state without, however, creating this state from a collision.
Of course, the notion of a collision in the $\nn=4$ theory is not
neccessarily very realistic as we have at our disposal only
quite specific configurations which may act as projectiles.

\begin{figure}[t]
\centerline{\includegraphics[height=5cm]{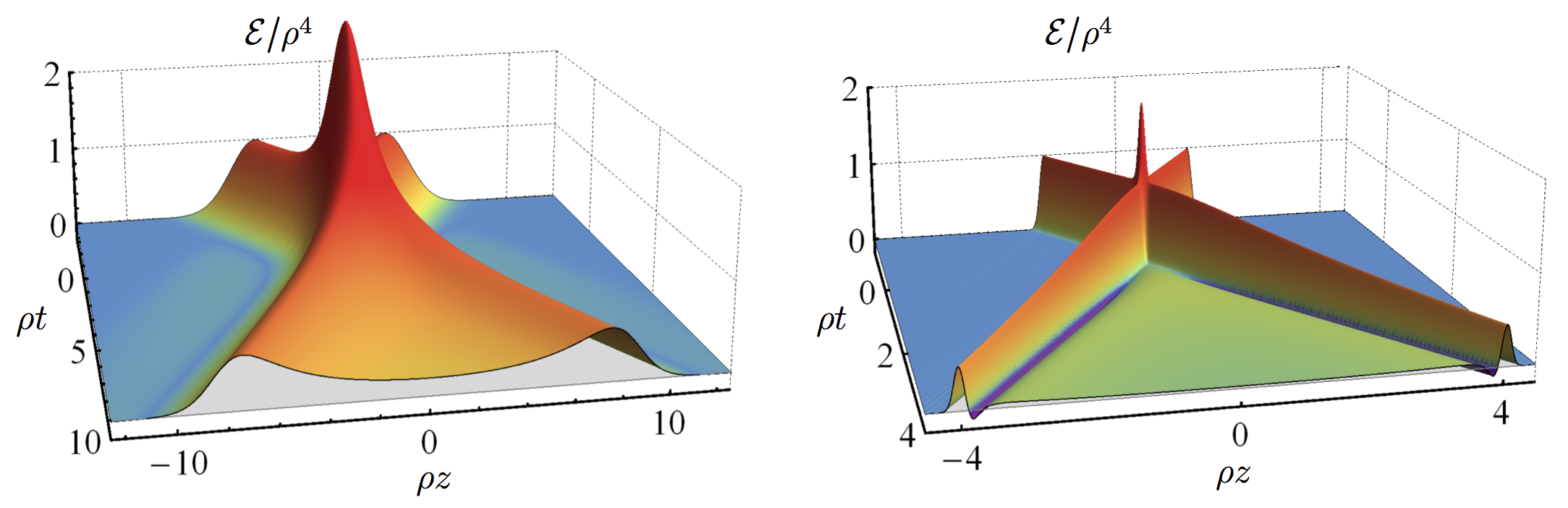}}
\caption{Energy profiles of thick and thin shock wave collisions (from \cite{MHSHOCK}).}
\label{fig4}
\end{figure}

A class of natural projectiles in the $\nn=4$ theory was suggested
in \cite{JP1}. These were shock waves corresponding to the energy momentum tensor
with the only nonvanishing component concentrated along one lightcone direction
\eq
\cor{T_{--}(x^-)}\propto \mu \dl(x^-) \quad\quad\text{or more generally}\quad 
\cor{T_{--}(x^-)}\propto f(x^-)
\eqx 
Their advantage is an explicit gravitational description
\eq
ds^2=\f{-dx^- dx^+ + f(x^-) z^4 {dx^-}^2+dx^2_\perp}{z^2}+\f{dz^2}{z^2}
\eqx
and the fact that they can be studied wholly within gravity without any excision (there are no
sources in the bulk in contrast to the shock waves employed e.g. in \cite{GUBSERSW}).
Unfortunately the gravitational description of a collision of two such shock waves
is still a very complicated problem.

The first few orders in a power expansion in $\tau$ of the geometry were obtained
in \cite{GR} in the case of thin shock waves. A first numerical investigation
of gaussian smeared shock waves was performed in \cite{CY}. In this case of
relatively thick shock waves, one found a scenario of full stopping (see \cite{MHSHOCK}
for this interpretation) and remnants
moving away with a speed $v<c$.

The very recent study \cite{MHSHOCK} revisited the issue of shock wave collision
and investigated shock waves of varying width (see Fig.~\ref{fig4}). 
It turned out that the resulting dynamics is surprisingly varied.
The chief outcome
was that the character of the collision significantly depends on the
thickness of the shock wave (namely the width in $x^-$ of the profile $f(x^-)$
relative to energy per
unit transverse area). For thick shock waves full stopping was recovered as in \cite{CY},
however for thin shock waves, fragments move away at speed of light with regions
of trailing negative energy density (reminiscent of some features of the analytical picture
of \cite{GR}). The paper \cite{MHSHOCK} also analyzed the detailed features of
the resulting hydrodynamic flow. One observation was a gaussian rapidity
profile. Thus there was no
restoration of strict boost-invariance. However it seems that a significant
part of the hydrodynamic flow can be well described by a (viscous) Bjorken flow with
the single scale $\Lm$ appearing in the Bjorken asymptotics ($\eps(\tau) \sim \Lm/\tau^{\f{4}{3}}$) being rapidity dependent $\Lm \to \Lm(y)$ \cite{Chesler}. 
However currently there does not seem to be a complete consensus on this issue.

\subsubsection*{Study V. A hybrid treatment of central nuclear collisions}

\begin{figure}[t]
\centerline{\includegraphics[height=5cm]{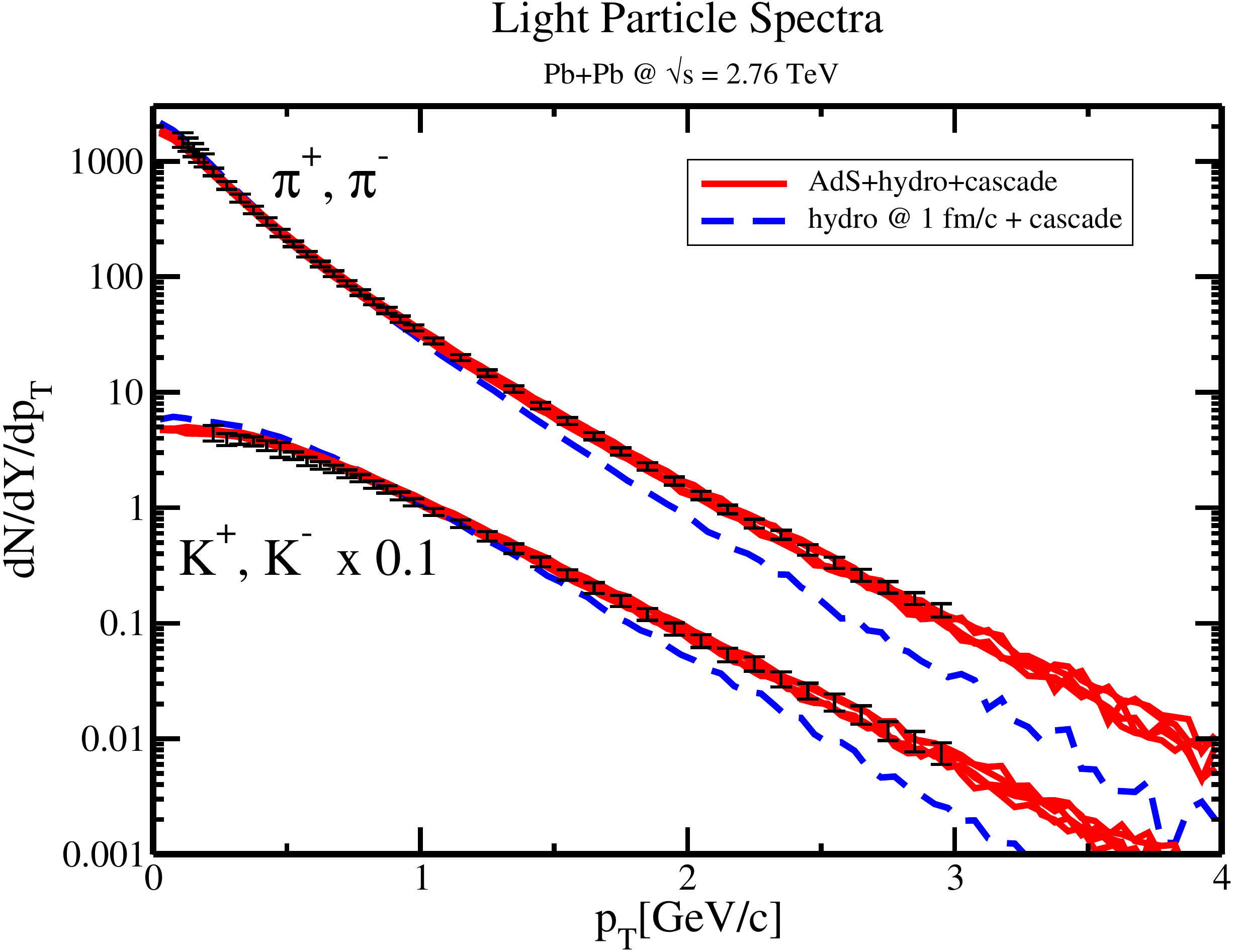}}
\caption{A comparision of experimental spectra of central \emph{Pb-Pb}
collisions from ALICE with i) the hybrid AdS/CFT+hydro code and
ii) hydro code with conventional initial conditions (from \cite{WRS}).}
\label{fig5}
\end{figure}

One of the features of the preceeding investigations was that they dealt directly
with strongly coupled plasma of the $\nn=4$ SYM theory. In a recent paper~\cite{WRS},
the authors adopted a very interesting hybrid approach using the AdS/CFT correspondence
(and hence the $\nn=4$ plasma) to model the initial nonequilibrium stage of the collision 
until the transition to a hydrodynamic description. Then from the resulting energy momentum
tensor they extracted the flow velocities and energy density/pressure and used
it as initial conditions for a standard realistic viscous hydrodynamic code coupled with a 
hadronic cascade model to describe the particles measured in the detector.

Let us note that such a matching between AdS/CFT and realistic hydro has some inherent
problems and ambiguities -- in particular the equation of state in the hydro code is
different from the one for $\nn=4$ plasma leading to some discontinuities in the pressure, however the authors found that their results
are quite stable and robust w.r.t. modifying the matching time between AdS/CFT and the hydro
simulation. The authors also adopted some approximations in the treatment of the
shock wave collision with radial profiles. The obtained results are, however, 
very promising as seen in Fig.~\ref{fig5}.

In this study, the novel input of the AdS/CFT correspondence was a prescription
for the initial conditions for hydrodynamics which naturally included
significant nontrivial radial flow already at the very begining of the applicability
of hydrodynamics. As emphasized earlier this is a very generic feature of
modelling the nonequilibrium phase using AdS/CFT methods.
It is very encouraging that the preequilibrium radial flow emerging
even from the approximate AdS/CFT treatment using the $\nn=4$ theory, 
goes in the direction of improving the description of light particle spectra.

\section{Conclusions}

The AdS/CFT correspondence provides a very general framework for studying
time-dependent dynamics of strongly coupled gauge theory plasma. What is
especially important is that it does not presuppose hydrodynamics
and is applicable even in very out-of-equillibrium contexts.
The various studies briefly reviewed in this talk provide examples
of qualitative insight that one may gain from these techniques
e.g. sizable pressure anisotropy at the transition to viscous
hydrodynamic description, a simple dimensionless estimate
of the `hydrodynamization' time, the importance of an `initial entropy'
as a characterization of initial conditions and subsequent dynamics,
the varied range of behaviours from full stopping to transparency
in shock wave collisions and finally the importance of preequilibrium 
radial flow.

\acknowledgments 
\noindent{}We thank the authors of \cite{MHSHOCK,WRS} for sharing the relevant figures.
This work was supported by NCN grant 2012/06/A/ST2/00396. RJ thanks IPhT Saclay for hospitality
when preparing this paper.

\end{document}